\documentclass[twocolumn,showpacs,amsmath,amssymb,nofootinbib]{revtex4}

\usepackage{graphicx}

\begin{document}


\title{Effects of Sequence Disorder on DNA Looping and Cyclization}

\author{Yuri O.\ Popov}
\email[Email address: ]{yopopov@mrl.ucsb.edu}
\affiliation{Materials Research Laboratory, University of California, Santa Barbara, California 93106, USA}
\author{Alexei V.\ Tkachenko}
\affiliation{Department of Physics, University of Michigan, 450 Church Street, Ann Arbor, Michigan 48109, USA}

\date{\today}

\begin{abstract}
Effects of sequence disorder on looping and cyclization of the double-stranded DNA are studied theoretically.  Both random intrinsic curvature and inhomogeneous bending rigidity are found to result in a remarkably wide distribution of cyclization probabilities.  For short DNA segments, the range of the distribution reaches several orders of magnitude for even completely random sequences.  The ensemble averaged values of the cyclization probability are also calculated, and the connection to the recent experiments is discussed.
\end{abstract}


\maketitle

\section{Introduction}

Bending of double-stranded (ds) DNA molecules into loops is important for such biological processes as regulation of gene expression and DNA packaging into nucleosomes~\cite{alberts,ptashne}.  Understanding the underlying physics of DNA looping is necessary for any quantitative description of these biological processes.  While the classical theory based on the elastic description of the DNA was proposed more than two decades ago~\cite{shimada}, there are multiple indications that it poorly describes the experimental situation.  In fact, the discrepancy between this theory and the experiment~\cite{bellomy,muller,cloutier1,cloutier2,du} often reaches several orders of magnitude.  Partially, this difference can be attributed to multiple complications arising from the \textit{in vivo} conditions.  However, recent experiments performed \textit{in vitro} still show substantial inadequacy of the existing theoretical description.  Several recent models attributed this discrepancy to the possibility of local disruptions of the ds-DNA structure such as ``kinks''~\cite{wiggins1} and ``bubbles''~\cite{yan,ranjith}.  However, both the classical description and the recent extensions ignore the sequence disorder, an intrinsic property of a typical DNA.

In this paper, we discuss the effect of sequence disorder on loop formation in DNA.  The inhomogeneity of the DNA structure is inherent to its biological function.  In nature, it is likely that sequences of the DNA segments involved in looping have evolved to optimize their bending properties.  Here we limit ourselves to the study of completely \textit{random} DNA sequences similar to those in recent experiments~\cite{cloutier1,cloutier2,du}.  Even in this simplest case, sequence-dependent effects lead to a remarkably wide distribution of the expected values of the looping probability and thus \textit{must} be taken into account in any realistic description.

\section{DNA looping problem}

The conformational properties of ds-DNA are commonly described by the worm-like chain (WLC) model, which proved to be particularly successful in accounting for the results of the single-molecule DNA stretching~\cite{marko}.  Within this model, the DNA is viewed as an inextensible elastic rode.  Let $L$ be the length of the chain, $s$ be the coordinate along the chain ($0 \le s \le L$), $\mathbf{r}(s)$ be the conformation of the chain, $\mathbf{t}(s) \equiv \partial\mathbf{r}/\partial s$ be the unit tangent vector at $s$, and $\mathbf{k}(s) \equiv \partial^2\mathbf{r}/\partial s^2$ be the chain curvature at $s$.  Then the effective Hamiltonian of the WLC model can be written as a sum of the bending and torsional energies:
\begin{equation}
\frac{H_{WLC}[\mathbf{k}]}{kT} = \frac{l_p}2 \int_0^L \left[ \mathbf{k}(s) \right]^2 ds + \text{torsion}.
\label{hamiltonian-wlc}
\end{equation}
Here and below, $l_p$ is the persistence length, a parameter proportional to the effective bending modulus and playing the role of the orientational correlation length of a free chain.  A similar parameter for the torsional contribution is known as the torsional persistence length $l_t$.  The experimental values of these parameters are $l_p \approx 50$~nm and $l_t \approx 70${\textendash}110~nm under physiological conditions.

The classical theory of DNA looping based on the WLC model was developed by Shimada and Yamakawa~\cite{shimada}.  Looping probability is proportional to the so-called $J$-factor, which is defined as the effective concentration of one of the chain ends in the vicinity of the other.  Formally, the $J$-factor is given as the canonical ensemble probability of the cyclized state:
\begin{equation}
J = \frac{\int D\mathbf{k} \, \exp \left( - \frac{H[\mathbf{k}]}{kT} \right) B[\mathbf{k}]}{\int D \mathbf{k} \, \exp \left( - \frac{H[\mathbf{k}]}{kT} \right)},
\label{j-definition}
\end{equation}
where the path integrations are over all possible conformations of the chain.  Function $B[\mathbf{k}]$ in the numerator represents a set boundary constraints in a given physical situation.  In particular, one can distinguish between protein-mediated looping (common for gene regulation) and cyclization (achieved by hybridization of the complementary single-stranded ends of a DNA segment).  The general case of DNA looping may involve rather complicated set of boundary conditions.  In this paper we focus on the problem with boundary constraints
\begin{equation}
B[\mathbf{k}] = \delta\left[\mathbf{r}(L) - \mathbf{r}(0)\right] \delta\left[\mathbf{t}(L) - \mathbf{t}(0)\right],
\label{bc}
\end{equation}
i.e.\ we assume alignment of both positions [$\mathbf{r}(L) = \mathbf{r}(0)$] and orientations [$\mathbf{t}(L) = \mathbf{t}(0)$] of the chain ends.  [Here $\mathbf{r}$ and $\mathbf{t}$ are understood as functionals of $\mathbf{k}(s)$.]  These boundary constraints are commonly employed to model DNA \textit{cyclization}.  In this case, the Shimada-Yamakawa result for $L < 4 l_p$ is given by~\cite{shimada}
\begin{equation}
J_{SY}(L) = \frac{C_1}{l_p^3} \left( \frac{l_p}L \right)^6 \exp\left[ - \frac{2 \pi^2 l_p}L + C_2 \frac{L}{l_p} \right] J_t \left(\frac{L}{l_t},\frac{L}{h}\right),
\label{shimada-yamakawa}
\end{equation}
where $C_1$ and $C_2$ are known numerical coefficients.  Factor $J_t$ is the contribution of the torsional constraints giving rise to the rapid oscillations of $J_{SY}(L)$ with the period of the helix repeat $h = 10.5$~bp.  For short DNA molecules, the overall result is dominated by the bending energy $E/kT = 2 \pi^2 l_p / L$ of the ground state (a circle).  The prefactor and the subdominant $C_2$-term are of entropic origin.  Modifications of the torsional contribution due to the sequence disorder are a smaller effect and will not be discussed in this paper.  Thus, $J_t$ will be assumed to have its classical form and omitted from all the expressions below for simplicity.

Cyclization properties of DNA molecules were studied experimentally by several groups~\cite{bellomy,muller,cloutier1,cloutier2,du}.  The $J$-factor is typically measured as the ratio of the equilibrium constants for chain cyclization and bimolecular association.  Substantial quantitative disagreement with the Shimada-Yamakawa theory was reported for short chain lengths, with experimental results showing significantly \textit{higher} values of the $J$-factor than allowed by the theory (up to several orders of magnitude).  In early works\cite{bellomy,muller} this disagreement was attributed to the difference between the \textit{in vivo} and \textit{in vitro} conditions.  However, the latest experiments~\cite{cloutier1,cloutier2,du} conducted \textit{in vitro} still indicate considerable deficiency of the classical theory.  Cloutier and Widom~\cite{cloutier1,cloutier2} reported cyclization probabilities four orders of magnitude higher than predicted theoretically.  At the same time, the results of Du \textit{et al.}~\cite{du} were different from those of Cloutier and Widom and closer to the classical prediction.  Nevertheless, their data could only be fitted by using the value of the persistence length as a fitting parameter.  Thus, the cyclization probability appears substantially underestimated by the classical theory for short DNA chains, and different groups report seemingly different experimental results.

Two effects of the sequence disorder are studied in this paper.  First, the local structure of the ds-DNA can be characterized by the sequence-dependent intrinsic curvature $\mathbf{k}_0(s)$.  Second, structural inhomogeneity naturally results in variations of the bending modulus along the chain.  This can be accounted for by the $s$-dependent persistence length $l_p(s)$.  The corresponding generalization of the WLC model is described by the following effective Hamiltonian:
\begin{equation}
\frac{H[\mathbf{k}]}{kT} = \frac{1}2 \int_0^L l_p(s) \left[ \mathbf{k}(s) - \mathbf{k}_0(s) \right]^2 ds + \text{torsion}.
\label{hamiltonian}
\end{equation}
Looping and cyclization occur on the mesoscopic scales, with characteristic lengths larger than a single base pair and comparable to the persistence length.  The above Hamiltonian is particularly suitable in this regime, where the elastic description is still valid, while the inhomogeneity of the elastic parameters is significant.  When $\mathbf{k}_0(s)$ is identically zero and $l_p(s)$ is a constant, this Hamiltonian reduces to the WLC Hamiltonian~(\ref{hamiltonian-wlc}).  We discuss the two sequence-disorder effects separately in the following two sections, and then combine them in the Discussion section.

\section{Effect of random intrinsic curvature}

First, we consider the effect of the sequence-dependent intrinsic curvature and use the effective Hamiltonian of Eq.~(\ref{hamiltonian}) with constant $l_p$.  Each DNA molecule is not a zero-thickness one-dimensional line; it has a finite cross-section instead.  In general, an arbitrary cross-section can be characterized by two principal axes with two different values of the intrinsic curvature along each of them.  An important point is that these principal directions are determined by the structure of the molecule and do not depend on the molecular conformation.  In particular, the two principal axes have generally no relation to the normal and the binormal of a particular conformation.  Thus, a DNA chain can be characterized by two pairs of the orthonormal vectors in the plain normal to the tangent: pair $\mathbf{e}_1$ and $\mathbf{e}_2$ is firmly affixed to the internal structure, and pair $\mathbf{n}$ and $\mathbf{b}$ is a function of its conformation.

Vector of the intrinsic curvature $\mathbf{k}_0$ can be decomposed along the principal axes at every point along the chain: $\mathbf{k}_0 = k_1 \mathbf{e}_1 + k_2 \mathbf{e}_2$.  Since looping and cyclization occur on scales much longer than the base pair length, the distribution of the intrinsic curvature $w[k_1,k_2]$ for a random sequence is Gaussian with zero average due to the Central Limit Theorem:
\begin{equation}
w[k_1,k_2] = \prod_{i=1}^2 \exp \left[ - \frac{\iint k_i(s) \delta(s - s') k_i(s') \, ds ds'}{2 \kappa} \right],
\label{w}
\end{equation}
i.e.\ the intrinsic curvature is delta-correlated along the chain: 
\begin{equation}
\langle k_i(s) k_j(s') \rangle_{\mathbf{k}_0} = \kappa \delta_{ij} \delta(s-s').
\end{equation}
Here and below, angular brackets with subscript $\mathbf{k}_0$ refer to the averaging with distribution~(\ref{w}):
\begin{equation}
\langle f[k_1,k_2] \rangle_{\mathbf{k}_0} = \frac{\iint Dk_1 Dk_2 \, w[k_1,k_2] f[k_1,k_2]}{\iint Dk_1 Dk_2 \, w[k_1,k_2]}.
\label{averaging}
\end{equation}
Note that the asymmetry of the cross-section averages out on the scale of a few base pairs due to the helical structure of ds-DNA, and thus we assume the value of parameter $\kappa$ to be the same for both principal components.  From the consensus-scale data of Gabrielian \textit{et al.} for various base-pair combinations~\cite{gabrielian} we found the value of parameter $\kappa \approx 0.13/l_p$ for several long ($ \ge 10000$~bp) random sequences by using their ``Bend It'' DNA tools server.  Note that a recent numerical study by Rappaport and Rabin~\cite{rapaport} also considered effects of the spontaneous sequence-dependent curvature; however, most of their results arose from the non-zero average intrinsic curvature assumed in that work.  In our model \textit{all} the effects are solely due to the random contribution and completely average out for a sufficiently long chain.

Remarkably, one can obtain a simple and general result for the $J$-factor averaged over the ensemble with the above statistics.  The unconstrained partition function in the denominator of Eq.~(\ref{j-definition}) is essentially a Gaussian integral, which is independent of the mean value of the Gaussian.  Thus, the denominator is independent of the intrinsic curvature, and hence the averaging over the disorder affects only the numerator.  Averaging over disorder $\mathbf{k}_0$ is interchangeable with path integration over $\mathbf{k}$, and only the Boltzmann factor depends on $\mathbf{k}_0$.  By averaging the Boltzmann factor in the numerator of Eq.~(\ref{j-definition}) according to Eq.~(\ref{averaging}), the original Boltzmann factor with a renormalized persistence length can be recovered:
\begin{equation}
\langle \exp\left( - \frac{H[\mathbf{k}]}{kT} \right) \rangle_{\mathbf{k}_0} = \frac{\exp\left[ - \frac{1}{2} \frac{l_p}{1 + \kappa l_p} \int_0^L \left[\mathbf{k}(s)\right]^2 ds \right]}{(1 + \kappa l_p)^{L/\xi}},
\label{boltzmann}
\end{equation}
\textit{independently} of the angle between $\mathbf{n}$ and $\mathbf{e}_1$ at every point along the chain.  Here, $\xi$ is the curvature correlation length of the order of a few base pairs.  Since the denominator of Eq.~(\ref{j-definition}) is a Gaussian integral, it scales with the persistence length as $l_p^{- L/\xi}$ (times an $l_p$-independent factor).  Thus, this denominator can be combined with $(1 + \kappa l_p)^{L/\xi}$ of expression~(\ref{boltzmann}) to recover the same denominator, but as a function of the renormalized persistence length: $[(1 + \kappa l_p)/l_p]^{L/\xi}$ (times the same $l_p$-independent factor).  Thus, the following \textit{exact} result for the cyclization probability can be obtained:
\begin{equation}
\langle J(L, l_p) \rangle_{\mathbf{k}_0} = J_{SY} (L, l_p^*),\quad\text{where}\quad l_p^* = \frac{l_p}{1 + \kappa l_p}
\label{j-curvature}
\end{equation}
is the \textit{renormalized} persistence length.  As a result of our averaging over the disorder, we recovered the original Shimada-Yamakawa expression, but with $l_p^*$ instead of $l_p$.  The renormalized persistence length $l_p^*$ is lower than the original one $l_p$, and thus the DNA molecule is softer and easier to bend.  We emphasize that this result is valid for \textit{any} boundary constraints $B[\mathbf{k}]$; it is not necessarily limited to looping or cyclization problems and hence very general.  Note that a necessary condition of this theorem is that the open configuration [described by the denominator of Eq.~(\ref{j-definition})] must be completely unconstrained, e.g.\ it \textit{cannot} be under an external force.  However, this theorem \textit{can} be applied to DNA stretching experiments.  In this case the stretched chain corresponds to the numerator of Eq.~(\ref{j-definition}) with boundary constraints $B[\mathbf{k}]$ representing the condition of the fixed end-to-end distance.  Thus, the same renormalization does take place in the stretching experiments, as was previously discussed in Refs.~\cite{trifonov,nelson,bensimon}.  Therefore, value of the persistence length extracted from such experiments ($\approx 50$~nm) is $l_p^*$ rather than $l_p$, and thus $\langle J \rangle_{\mathbf{k}_0}$ actually \textit{coincides} with the Shimada-Yamakawa result, plotted for the \textit{experimental} values of this parameter.

However, $\langle J \rangle_{\mathbf{k}_0}$ is not the only characteristics of the cyclization probability for random sequences, and the width of the distribution of $J$ turns out to be equally important.  The latter can be deduced from the comparison of $\log \langle J \rangle_{\mathbf{k}_0}$ with $\langle \log J \rangle_{\mathbf{k}_0}$.  Moreover, if the distribution of $\log J$ is Gaussian, then the average coincides with the maximum of probability and hence $\langle \log J \rangle_{\mathbf{k}_0}$ is the value most likely to be observed when plotting the experimental data on the logarithmic scale.  Finally, $\langle \log J \rangle_{\mathbf{k}_0}$ has a clear physical meaning of (the negative of) the average free energy difference between the looped and the unconstrained states.  

\begin{figure}
\includegraphics{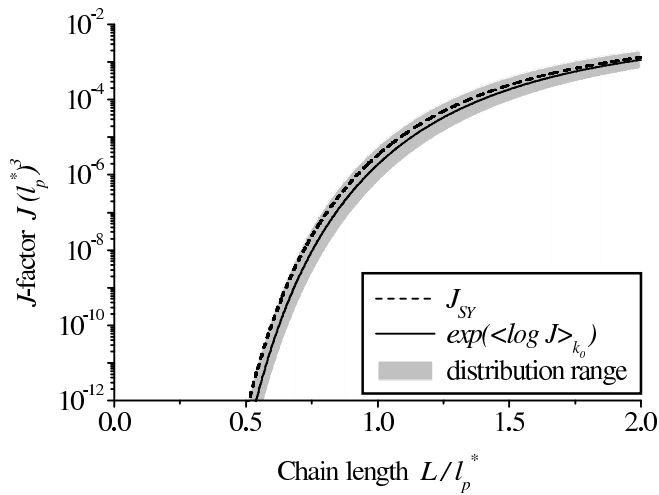}
\caption{\label{curvature} Effect of the intrinsic curvature on the cyclization probability.  The contribution of the torsional constraints to the $J$-factor is omitted.  The dashed line represents the average value of the $J$-factor over the intrinsic curvature [Eq.~(\ref{j-curvature})], which coincides with the Shimada-Yamakawa result for the experimental values of the persistence length.  The solid line is the average value of $\log J$, i.e.\ the interpolation between Eq.~(\ref{log-j-curvature}) for short chains and $\log \langle J \rangle_{\mathbf{k}_0}$ for long chains.  The width of the distribution is indicated by the grey shading; it reaches an order of magnitude for $L = l_p^*/2$.  Two thirds of random sequences are supposed to lie within the shaded area.}
\end{figure}

Quantity $\langle \log J \rangle_{\mathbf{k}_0}$ can be calculated in a fashion similar to $\langle J \rangle_{\mathbf{k}_0}$.  For short chains, the average free energy difference between the looped and the unconstrained states is dominated by the average difference in the energy of the ground state.  The energy of the ground state is given by the Hamiltonian minimized with respect to the angle between $\mathbf{n}$ and $\mathbf{e}_1$.  Thus, conducting the averaging over the disorder of the minimized Hamiltonian, one can obtain the following result for $\langle \log J \rangle_{\mathbf{k}_0}$ (for short enough chains): 
\begin{equation}
\langle \log J \rangle_{\mathbf{k}_0} = - \frac{2 \pi^2 l_p}L + \pi l_p \sqrt{\frac{2 \pi \kappa}L} + O \left(\log \frac{L}{l_p}\right).
\label{log-j-curvature}
\end{equation}
The first (dominant) term is the bending energy of the circle as in the Shimada-Yamakawa result.  The second term originates from the cross-term in the Hamiltonian and reflects the fact that a \textit{finite-length} segment has a non-zero mean intrinsic curvature of the order of $\sqrt{\kappa / L}$ despite the ensemble-average curvature being zero.  This correction raises the expectation value of $\log J$ (although the result is still lower than $\log \langle J \rangle_{\mathbf{k}_0}$ for short chains, as expected).  For longer chains $\langle \log J \rangle_{\mathbf{k}_0} = \log \langle J \rangle_{\mathbf{k}_0}$ due to self-averaging.  This result for long chains and result~(\ref{log-j-curvature}) for short chains can be interpolated to the entire range of the chain lengths.  The result of this interpolation (which is asymptotically correct in both limits) is plotted in Fig.~\ref{curvature} together with the sequence-averaged $\langle J(L) \rangle_{\mathbf{k}_0}$ of Eq.~(\ref{j-curvature}) (which coincides with the Shimada-Yamakawa result).

For weak enough disorder the free energy and therefore $\log J$ of the cyclized state have a Gaussian distribution.  In this case, its width $\delta$ can be inferred from the two averages $\delta^2 = 2 \left( \log \langle J \rangle_{\mathbf{k}_0} - \langle \log J \rangle_{\mathbf{k}_0} \right)$ and is also shown in Fig.~\ref{curvature}.  This width is substantial, and the distribution spans an order of magnitude for short chains.

\section{Effect of inhomogeneous bending rigidity}

The second effect of the sequence disorder deals with the inhomogeneous bending rigidity and can be described by the effective Hamiltonian of Eq.~(\ref{hamiltonian}) with zero intrinsic curvature $\mathbf{k}_0$ but $s$-dependent persistence length $l_p$.  We limit our discussion to the case of the isotropic bending [described by a \textit{single} local bending modulus $k T l_p(s)$], although a more general exposition should allow for the anisotropy of the local bending modulus as well.

The major effect is related to the correction to the energy of the ground state.  Consider a chain segment of length $L$ bent by a small angle $\Delta \theta$.  Since the intrinsic curvature is zero at all points, the chain segment is confined to a plain and can be described by angle $\theta(s)$ in that plain.  Torque $l_p(s) \partial\theta/\partial s$ is constant along the chain and equal to $\Delta \theta / \int_0^L l_p^{-1}(s) ds$.  Hence the bending energy can be written as
\[
\frac{E}{kT} = \int_0^L \frac{l_p(s)}2 \left( \frac{\partial\theta}{\partial s} \right)^2 ds = \int_0^L \frac{l_p(s) \left( \Delta \theta \right)^2 ds}{2 \left[ l_p(s) \int_0^L l_p^{-1}(s') ds' \right]^2}
\]
or
\begin{equation}
\frac{E}{kT} = \frac{(\Delta \theta)^2 l_p}{2 x L},\quad\text{where}\quad x = \frac{l_p}L \int_0^L \frac{ds}{l_p(s)}
\label{energy-rigidity}
\end{equation}
is a dimensionless parameter.  Here and below we distinguish between the ensemble-wide harmonic average persistence length $l_p \equiv \langle l_p^{-1} \rangle^{-1}$ and the local $s$-dependent value $l_p(s)$.  Equation~(\ref{energy-rigidity}) implies that the chain segment can be described by the \textit{effective} bending modulus $k T l_p / x$ instead of $k T l_p$.

In full analogy with the case of the intrinsic curvature, it is the \textit{harmonic} average that is relevant to both DNA cyclization/looping and DNA stretching problems, as was first pointed out by Wiggins~\cite{wiggins2}.  This observation is valid when the correlation length of the disorder is much smaller than the persistence length, which is indeed a typical situation for DNA.  Mathematically, the origin of the harmonic average is the same as in the problem of several Hookean springs connected in series, where the effective spring constant for the series is the harmonic average of the respective spring constants.

The concept of the effective bending modulus can now be applied to the cyclization problem by imposing the constraint $\mathbf{t}(L) = \mathbf{t}(0)$, i.e.\ assuming $\Delta \theta = 2 \pi$ in Eq.~(\ref{energy-rigidity}).  Strictly speaking, the other constraint $\mathbf{r}(L) = \mathbf{r}(0)$ is not automatically satisfied as in the uniform case.  However, the corresponding correction to the overall bending energy is weaker than the effect of the renormalization of the bending modulus and hence neglected (see Appendix for details).  We also ignore any corrections to the subdominant (for $L < l_p$) entropic terms in the Shimada-Yamakawa result.  Thus, the only modification of the expression for $J$ is in the ground state bending energy:
\begin{equation}
J(L) = J_{SY}(L) \exp \left[ - \frac{2 \pi^2 l_p}{L} \left(\frac{1}x - 1\right) \right].
\label{modification}
\end{equation}
 
Now, all statistical properties of the $J$-factor are completely determined by those of quantity $x$.  Random quantity $l_p^{-1}(s)$ in the definition of $x$ is correlated on the length scale $\Delta l$ of the order of a few base pairs, which is much shorter than $L$.  Thus, according to the Central Limit Theorem, the distribution of $x$ is \textit{Gaussian}.  Parameters of this distribution are fully defined by the average and the second-order correlator of $l_p^{-1}(s)$.  By definition, $\langle l_p^{-1}(s) \rangle \equiv l_p^{-1}$, therefore $\langle x \rangle = 1$.  Since $\Delta l \ll L$, the correlator of $l_p^{-1}(s)$ can be written as
\begin{equation}
\langle \left[ l_p^{-1}(s) - l_p^{-1} \right] \left[ l_p^{-1}(s') - l_p^{-1} \right] \rangle = \frac{\alpha}{l_p} \delta(s-s').
\label{lps}
\end{equation}
Here $\alpha$ is a dimensionless parameter related to the correlation length $\Delta l$ and the second moment of the local distribution of $l_p^{-1}(s)$:
\begin{equation}
\alpha = l_p \Delta l \langle \left[ l_p^{-1}(s) - l_p^{-1} \right]^2 \rangle.
\end{equation}
This parameter can be estimated from a set of the consensus-scale data of Munteanu \textit{et al.}~\cite{munteanu}.  Their data lists sequence-dependent Young's modulus for all 64 possible trinucleotides.  We find that $\langle \left[l_p^{-1}(s) - l_p^{-1} \right]^2 \rangle \approx 0.13 l_p^{-2}$.  Assuming $\Delta l \approx 3\text{ bp}$ and $l_p \approx 150\text{ bp}$, we find $\alpha \approx 0.13 \Delta l/ l_p \approx 0.0025$.  Note that a trinucleotid may be too approximate a candidate for the correlation length $\Delta l$, and hence the above value of coefficient $\alpha$ is just an order-of-magnitude estimate.  From Eq.~(\ref{lps}) we can obtain the dispersion of $x$:
\[
\sigma^2 \equiv \langle \left( x - \langle x \rangle \right)^2 \rangle 
\]
\begin{equation}
= \frac{l_p^2}{L^2} \iint_0^L \langle \left[ l_p^{-1}(s) - l_p^{-1} \right] \left[ l_p^{-1}(s') - l_p^{-1} \right] \rangle ds ds' = \frac{\alpha l_p}L.
\end{equation}
Thus, the overall (normalized) distribution of $x$ is:
\begin{equation}
\phi(x) = \sqrt{\frac{L}{2\pi\alpha l_p}} \exp\left[ - \frac{L}{2 \alpha l_p} \left( x - 1 \right)^2 \right].
\label{gaussian}
\end{equation}
Note that we did not assume the statistics of $l_p^{-1}(s)$ to be Gaussian in this derivation.

\begin{figure}
\includegraphics{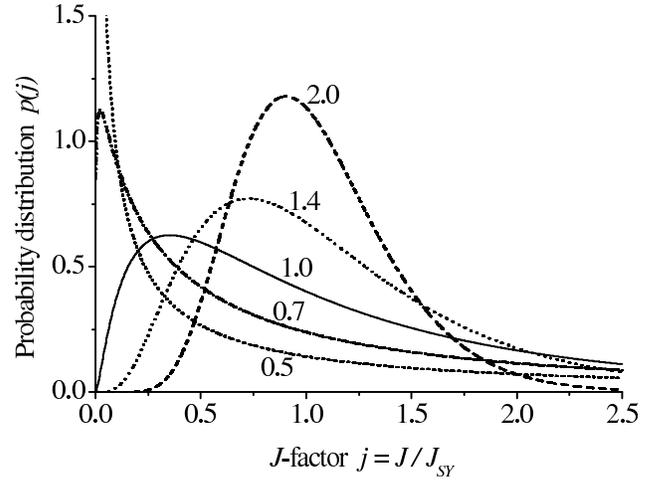}
\caption{\label{distribution} Probability distribution~(\ref{probability}) of the $J$-factor~(\ref{modification}) for various values of the chain length: $L \langle l_p^{-1} \rangle = 2.0$, 1.4, 1.0, 0.7, and 0.5.  As the chain length decreases, the distribution evolves from nearly Gaussian to significantly non-Gaussian, with a strong tail developing at large values of $J$.  The contribution of these large but rare values raises the average value of the $J$-factor substantially.}
\end{figure}

Based on the statistics of $x$, one can obtain the distribution of $J$ of Eq.~(\ref{modification}):
\begin{equation}
p(j) = \frac{\sqrt{2 \pi^3 \alpha^{-1} l^3}}{j \left(2 \pi^2 - l \log j \right)^2} \exp\left[ - \frac{l}{2 \alpha} \left(\frac{l \log j}{2 \pi^2 - l \log j}\right)^2 \right],
\label{probability}
\end{equation}
where $j \equiv J / J_{SY}$ and $l \equiv L / l_p$.  This distribution is plotted in Fig.~\ref{distribution} for several values of $L$.  It turns out to be quite peculiar: one can observe a striking crossover from a nearly Gaussian shape to a significantly non-Gaussian one as the chain length decreases.  For short chains the distribution behaves as a power law $1/J$ over an increasing range of values of $J$, giving rise to a very long tail.  As a result, the average $\langle J \rangle_{l_p}$ is dominated by large but rare values.  Here and below, angular brackets with subscript $l_p$ refer to the averaging with distribution~(\ref{gaussian}):
\begin{equation}
\langle f(x) \rangle_{l_p} = \int f(x) \phi(x) dx,
\end{equation}
or, equivalently, $\langle f(j) \rangle_{l_p} = \int f(j) p(j) dj$.

\begin{figure}
\includegraphics{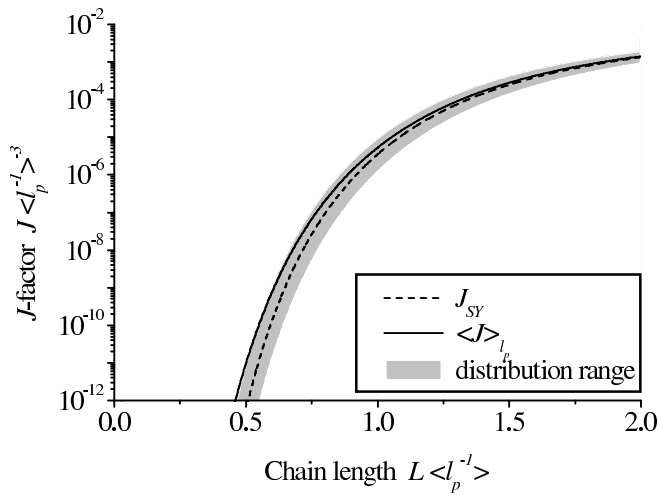}
\caption{\label{rigidity} Effect of the sequence-dependent bending rigidity on the cyclization probability.  The contribution of the torsional constraints to the $J$-factor is omitted.  The dashed line represents the Shimada-Yamakawa result.  The solid line is the average value of the $J$-factor over the bending rigidity, which is strongly enhanced for short chains.  The width of the distribution is indicated by the grey shading; it exceeds an order of magnitude for short chains (and almost vanishes for long chains).}
\end{figure}

An analytical expression for $\langle J \rangle_{l_p}$ can be obtained from the distribution.  Conducting an integration by the method of the steepest descent, one can obtain the average of the leading factor $\langle \exp \left(-E/kT\right) \rangle_{l_p}$: 
\begin{equation}
\langle \exp \left( - \frac{E}{kT} \right) \rangle_{l_p} = \frac{\exp\left[ - \frac{2 \pi^2 l_p}{L x_0} - \frac{L (x_0 - 1)^2}{2 \alpha l_p} \right]}{\sqrt{1 + \frac{\alpha}{x_0^3} \left(\frac{2 \pi l_p}{L}\right)^2}},
\end{equation}
where $x_0$ is the saddle point:
\begin{equation}
x_0 = \frac{1}{3} \left( \gamma + 1 + \frac{1}{\gamma} \right),
\end{equation}
\[
\gamma = \left( \delta + 1 + \sqrt{\delta^2 + 2\delta} \right)^{1/3},\quad\text{and}\quad \delta = 27 \alpha \left(\frac{\pi l_p}{L}\right)^2.
\]
For short chains [shorter than $(27 \alpha)^{1/2} \pi l_p$, or about $0.82 l_p$], $\delta \gg 1$ and $x_0 \approx (2 \delta)^{1/3}/3$.  Thus, the short-length behavior of this leading factor is dramatically different in its functional form from $\exp \left( - 2 \pi^2 l_p / L \right) $ of the Shimada-Yamakawa result:
\begin{equation}
\langle \exp \left( - \frac{E}{kT} \right) \rangle_{l_p} \propto \frac{1}{\sqrt{3}} \exp \left[ - \left( \frac{27 \pi^4}{2 \alpha} \frac{l_p}L \right)^{1/3} \right].
\label{j-rigidity}
\end{equation}
The full expression for $\langle J(L) \rangle_{l_p}$ is plotted in Fig.~\ref{rigidity} together with $J_{SY}(L)$.  The average $\langle J \rangle_{l_p}$ significantly exceeds the classical value $J_{SY}$ for short chains.

The characteristic range of the distribution of $J$ is indicated in Fig.~\ref{rigidity} by shading the area between the graphs corresponding to $x = 1 + \sigma$ and $x = 1 - \sigma$ in expression~(\ref{modification}).  This shaded area includes 68\% of random sequences since the distribution of $x$ is Gaussian and $J(x)$ is a monotonic function of $x$.  For short chains the widening of the probability distribution due to the inhomogeneous bending rigidity is even stronger than due to the intrinsic curvature.

\section{Discussion}

\begin{figure}
\includegraphics{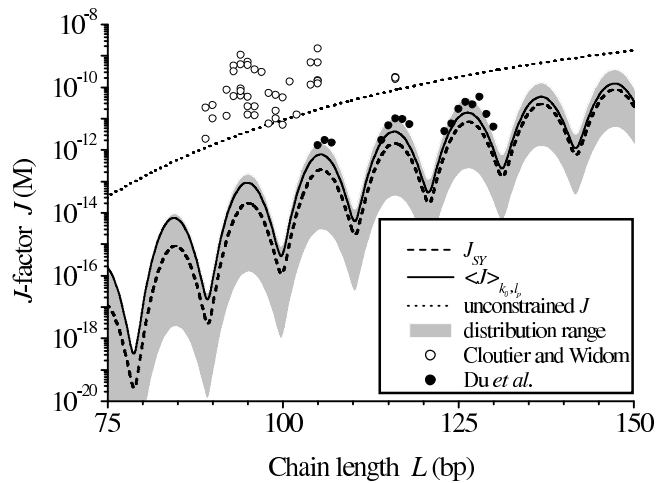}
\caption{\label{final} Overall effect of the sequence disorder on the cyclization probability.  The torsional contribution to the $J$-factor is taken into account, and physical units are restored.  The dashed line is the classical Shimada-Yamakwa result.  The average of the $J$-factor over both types of sequence disorder is represented by the solid line.  The distribution range is indicated by the grey shading; it spans three to four orders of magnitude for chains shorter than $2/3$ of the persistence length.  Comparison to the experimental data of Cloutier and Widom~\cite{cloutier1,cloutier2} and Du \textit{et al.}~\cite{du} is made.  While the absolute values of the experimental data of Cloutier and Widom are not reached for 68\% of random sequences, the spread of their data is consistent with our prediction.  The dotted line represents the $J$-factor for a loop with unconstrained orientations of the ends.  It indicates that the boundary conditions may be the clue to the discrepancy in absolute values.}
\end{figure}

Here we combine the contributions of both random intrinsic curvature and inhomogeneous bending rigidity.  In Fig.~\ref{final}, we present the overall result and restore torsional effects giving rise to the oscillatory behavior of $J(L)$.  The theoretical curves are plotted in real physical units and compared to experimental data.  The value of $J$ averaged over both types of disorder $\langle J(L) \rangle_{\mathbf{k}_0,l_p}$ and the typical range of its distribution are shown as functions of the chain length.

The discrepancy between the theory and the experimental data of Refs.~\cite{cloutier1,cloutier2} remains statistically significant.  While this deviation might be attributed to the effects of ``kinks''~\cite{wiggins1} and ``bubbles''~\cite{yan,ranjith}, we note that the overall theoretical result strongly depends on the choice of the boundary conditions.  The current experiments do not ensure that the ends of the looped segment are perfectly aligned [$\mathbf{t}(L) = \mathbf{t}(0)$].  Rather, the cyclization is achieved by the hybridization of the complementary single-stranded ends.  Upon this process (and before the ligation) each of the two strands is not structurally continuous since there is no covalent bonding involved.  As a result, the flexibility on both sides of the newly hybridized region is higher than elsewhere along the double-stranded loop and could be modeled with modified boundary conditions.  The extreme case of a completely free joint at the location of the newly hybridized region has already been solved by Shimada and Yamakawa in Ref.~\cite{shimada}.  The result of this calculation (also shown in Fig.~\ref{final} by the dotted curve) is in better agreement with the experimental data of Refs.~\cite{cloutier1,cloutier2}.  A detailed analysis of a modified model with intermediate boundary conditions (neither completely free nor perfectly aligned tangent vectors) is conducted in Ref.~\cite{tkachenko}.

The overall width of the distribution reaches three orders of magnitude in the range of lengths of Fig.~\ref{final}.  This predicted width is consistent with the spread of the experimental data of Cloutier and Widom~\cite{cloutier1,cloutier2}.  On the other hand, we cannot make a conclusive comparison of the calculated width with the experimental data of Refs.~\cite{du,vologodskaia} due to insufficient statistics.

It should be emphasized that both the classical theory by Shimada and Yamakawa and its derivatives implicitly assumed the property of \textit{self-averaging}: the looping probability ($J$-factor) of a ``typical'' DNA segment was expected to follow a certain function of its length $L$, at least in the first approximation.  The major result of our work is that there is \textit{no} self-averaging in the problem of looping and cyclization in the most interesting regime of short chains ($L \lesssim l_p$).  The $J$-factor is \textit{not} a well-defined function of the length, not even to the first approximation.  In other words, there is no ``typical'' DNA segment in that regime.  The lack of self-averaging is indicated by extremely wide range of the distribution of the $J$-factors and by strongly non-Gaussian (power-law-like) behavior of that distribution.  This constitutes a drastic modification to the original theory (and any of its derivatives) and has far reaching implications for the future studies of the problem.  This also makes DNA looping and cyclization dramatically different from the traditional DNA stretching problem, where the property of self-averaging is automatically satisfied.  The notion of the probability of looping of short DNA segments is simply meaningless outside the context of a particular DNA sequence.

Due to this lack of self-averaging, future experiments should involve a sufficiently large ensemble of sequences of the same length in order to be conclusive.  For instance, by measuring the cyclization probability in an unbiased mixture of such DNA segments, one can find $\langle J \rangle$, which is expected to follow a functional form very different from the Shimada-Yamakawa result.  Similarly, the distribution function of the $J$-factor can be related to the statistics of the loop closing times.  The latter could be measured e.g.\ by using the molecular-beacon-type technique~\cite{molbec}.

The numerical values of parameters $\kappa$ and $\alpha$ may be a possible source of uncertainty in our results.  These values were based on the experimental data of Refs.~\cite{gabrielian} and \cite{munteanu}, which appears to be quite unreliable.  In particular, the difference between the reported sets of experimental values determined on the basis of two different models (``Consensus'' and ``DNasel'') is rather substantial and reaches a factor of two or three for certain trinucleotides~\cite{munteanu}.  Given the exponential nature of the dependence of the $J$-factor on parameters $\kappa$ and $\alpha$, one can expect a major numerical variation in our final results if a different experimental data set is employed.  When we used the DNasel values instead of the Consensus ones, we did observe almost an order of magnitude variation of the $J$-factor for short chains.  In the future, we expect that more reliable experimental data will become available and more accurate estimates of $\kappa$ and $\alpha$ can be obtained.

Our results suggest that in nature DNA looping may have been substantially optimized by the evolutionary selection of the appropriate sequences.  By comparing the experiments on DNA looping and cyclization performed over natural sequences to our results for generic random sequences, one can estimate the significance of this evolutionary optimization.

\begin{acknowledgments}
The authors are grateful to P.~A.~Wiggins for indispensable comments on the early version of this paper, to J.-C.~Meiners and Y.~Rabin for valuable discussions, and to J.~Widom and A.~Vologodskii for providing experimental data.
\end{acknowledgments}

\appendix*
\section{}

The purpose of this appendix is to demonstrate that the additional boundary constraint $\mathbf{r}(L) = \mathbf{r}(0)$ contributes a small correction to the bending energy of the loop, Eq.~(\ref{energy-rigidity}), obtained under boundary conditions $\mathbf{t}(L) = \mathbf{t}(0)$ only.

Consider a looped DNA chain.  Since the intrinsic curvature is assumed to be zero, the ground state of such a chain is confined to a plane.  Let $\mathbf{b}$ be a unit vector normal to that plane and $\theta(s)$ be the in-plane orientation angle of a chain segment at coordinate $s$.  Let also $\mathbf{f}$ be the force acting between the two ends and $\tau$ be the constant torque along the chain; both are needed to enforce boundary conditions~(\ref{bc}).  The shape of the loop ground state is determined by the following equations:
\begin{equation}
\frac{\partial \mathbf{t}}{\partial s} = \frac{\partial\theta}{\partial s} \mathbf{b} \times \mathbf{t},
\end{equation}
\begin{equation}
\frac{\partial\theta}{\partial s} = \frac{\tau + \mathbf{b} \cdot \left( \mathbf{r}(s) \times \mathbf{f} \right)}{l_p(s)} = \frac{\tau + \left( \mathbf{r}(s) \cdot \mathbf{f}^* \right)}{l_p(s)},
\end{equation}
where we introduced $\mathbf{f}^* \equiv \mathbf{f} \times \mathbf{b}$.  The ground state bending energy can be formally expressed as a quadratic form with respect to $\tau$ and $\mathbf{f}^*$: 
\[
\frac{E}{kT} = \int_0^L \frac{l_p(s)}2 \left( \frac{\partial\theta}{\partial s} \right)^2 ds
\]
\begin{equation}
= \int_0^L \frac{\left[ \tau + \left( \mathbf{r}(s) \cdot \mathbf{f}^* \right) \right]^2}{2 l_p(s)} ds \equiv \frac{\vec{v}^{T} \cdot \hat{M} \cdot \vec{v}}{2}.
\label{En}
\end{equation}
Here we combined the torque and the force into a single ``vector'' $\vec{v}^{T} = \left(\tau, \mathbf{f}^*\right)$ and introduced ``matrix'' $\hat{M}$ that depends on the loop shape:
\begin{equation}
\hat{M} = \int_0^L \left(
\begin{array}{ccc}
1 & X & Y \\ 
X & X^2 & XY \\ 
Y & XY & Y^2
\end{array}
\right) \frac{ds}{l_p(s)}.
\end{equation}
In the last equation, $X(s)$ and $Y(s)$ are the in-plane components of the chain conformation $\mathbf{r}(s)$.  They should not be confused with ``components'' of ``vector'' $\vec{v}$, which we will distinguish by indices 1, 2, and 3 below.  ``Component'' 1 of that ``vector'' corresponds to the torque, and ``components'' 2 and 3 correspond to the two in-plane components of the ``force'' $\mathbf{f}^*$.

Now we need to choose the torque and the force in such a way that boundary conditions are satisfied.  In particular, the tangential constraint imposes: 
\begin{equation}
\int_0^L \frac{\partial\theta}{\partial s} ds = \frac{1}{kT} \frac{\partial E}{\partial\tau} = \left( \hat{M} \cdot \vec{v} \right)_1 = 2\pi.
\end{equation}
The other constraint, $\mathbf{u} \equiv \mathbf{r}(L) - \mathbf{r}(0) = 0$, is automatically satisfied in the uniform case [$l_p(s) = const$], when the ground state of the loop is a circle $\partial\theta/\partial s = 2\pi/L$.  In a moderately non-uniform case, one can express $\mathbf{u}$ in terms of small deviations of $\partial\theta/\partial s$ from its average value $2\pi/L$: 
\[
\mathbf{u} = \int_0^L \left( \frac{\partial\theta}{\partial s} - \frac{2\pi}{L} \right) \mathbf{b} \times \mathbf{r}(s) ds 
\]
\begin{equation}
= \mathbf{b} \times \left( \frac{1}{kT} \frac{\partial E}{\partial\mathbf{f}^*} - \frac{2\pi}{L} \int_0^L \mathbf{r}(s) ds \right) = 0.
\end{equation}
For convenience, we choose the origin to be in the center of mass of the loop.  In this case $\int_0^L \mathbf{r}(s) ds = 0$ and the last condition reduces to
\begin{equation}
\frac{1}{kT} \frac{\partial E}{\partial\mathbf{f}^*} = \left( \hat{M} \cdot \vec{v} \right)_{2,3} = 0.
\end{equation}

We can now calculate the overall energy by substituting $\vec{v}^{T} = \left( 2\pi, 0, 0 \right) \cdot \hat{M}^{-1}$ into Eq.~(\ref{En}): 
\[
\frac{E}{kT} = \frac{\left( 2 \pi \right)^2}{2} \left( \hat{M}^{-1} \right)_{11}
\]
\begin{equation}
= \frac{2 \pi^2 l_p}{x L} \left( 1 + \frac{2 \overline{X}\,\overline{Y}\,\overline{XY} - \overline{X}^2\,\overline{Y^2} - \overline{X^2}\,\overline{Y}^2}{\overline{X^2}\,\overline{Y^2} - \overline{XY}^2} \right)^{-1}
\label{e-m}
\end{equation}
Here the bar denotes averaging for a given sequence with a weight factor proportional to $1/l_p(s)$, for example:
\begin{equation}
\overline{X} = \frac{l_p}{x L} \int_0^L \frac{X ds}{l_p(s)};
\end{equation}
parameter $x$ is defined in Eq.~(\ref{energy-rigidity}).  Note that in the uniform case $\overline{X} = \overline{Y} = \overline{XY} = 0$ and $\overline{X^2} = \overline{Y^2} = L^2 / 8 \pi^2$.  Retaining only the leading terms in disorder, one can obtain from Eq.~(\ref{e-m}):
\begin{equation}
\frac{E}{kT} \approx \frac{2 \pi^2 l_p}{xL} \left( 1 + 8\pi^2 \frac{\overline{X}^2 + \overline{Y}^2}{L^2} \right).
\end{equation}
Introducing
\begin{equation}
w = \overline{X + i Y} \approx \frac{l_p}{x L} \int_0^L \frac{L}{2 \pi l_p(s)} \exp\left[ \frac{2 \pi i s}{L} \right] ds, 
\end{equation}
one can re-write the previous result as
\begin{equation}
\frac{E}{kT} \approx \frac{2 \pi^2 l_p}{xL} \left( 1 + \Delta \right),
\end{equation}
where $\Delta \equiv 8 \pi^2 w w^* / L^2$ and $w^*$ is the complex conjugate of $w$.  The dominant term in the last expression, $2 \pi^2 l_p / x L$, coincides with our result~(\ref{energy-rigidity}) obtained \textit{without} accounting for the additional constraint $\mathbf{r}(L) = \mathbf{r}(0)$.  Thus, constraining positions of the ends in addition to constraining their orientations adds only a correctional term to the bending energy of the ground state.  The ensemble average of correction $\Delta$ is given by:
\begin{equation}
\langle \Delta \rangle = \frac{8 \pi^2}{L^2} \langle w w^* \rangle \approx \frac{2 \alpha l_p}L.
\end{equation}
Since $\alpha \approx 0.0025$, correction $\langle \Delta \rangle$ is less than 1\% for all chain lengths $L > l_p / 2$.


\begin{thebibliography}{99}

\bibitem{alberts} B.~Alberts \textit{et al.}, \textit{Molecular Biology of the Cell} (Garland, New York, 1989).

\bibitem{ptashne} M.~Ptashne, \textit{A Genetic Switch} (Cell Press \& Blackwell Science, Cambridge, MA, 1992).

\bibitem{shimada} J.~Shimada and H.~Yamakawa, Macromolecules \textbf{17}, 689 (1984).

\bibitem{bellomy} G.~R.~Bellomy, M.~C.~Mossing, and M.~T.~Record,~Jr., Biochemistry \textbf{27}, 3900 (1988).

\bibitem{muller} J.~Muller, S.~Oehler, and B.~Muller-Hill, J.\ Mol.\ Biol.\ \textbf{257}, 21 (1996).

\bibitem{cloutier1} T.~E.~Cloutier and J.~Widom, Mol.\ Cell \textbf{14}, 355 (2004).

\bibitem{cloutier2} T.~E.~Cloutier and J.~Widom, Proc.\ Natl.\ Acad.\ Sci.\ U.S.A.\ \textbf{102}, 3645 (2005).

\bibitem{du} Q.~Du, C.~Smith, N.~Shiffeldrim, M.~Vologodskaia, and A.~Vologodskii, Proc.\ Natl.\ Acad.\ Sci.\ U.S.A.\ \textbf{102}, 5397 (2005).

\bibitem{wiggins1} P.~A.~Wiggins, R.~Phillips, and P.~C.~Nelson, Phys.\ Rev.\ E \textbf{71}, 021909 (2005).

\bibitem{yan} J.~Yan and J.~F.~Marko, Phys.\ Rev.\ Lett.\ \textbf{93}, 108108 (2004).

\bibitem{ranjith} P.~Ranjith, P.~B.~Sunil~Kumar, and G.~I.~Menon, Phys.\ Rev.\ Lett.\ \textbf{94}, 138102 (2005).

\bibitem{marko} J.~F.~Marko and E.~G.~Siggia, Macromolecules \textbf{28}, 8759 (1995).

\bibitem{gabrielian} A.~Gabrielian, K.~Vlahovicek, and S.~Pongor, DNA tools, \url{http://hydra.icgeb.trieste.it/~kristian/dna/}

\bibitem{rapaport} S.~Rappaport and Y.~Rabin, Macromolecules \textbf{37}, 7847 (2004).

\bibitem{trifonov} E.~N.~Trifonov, R.~K.-Z.~Tan, and S.~C.~Harvey, in \textit{DNA bending and curvature}, edited by W.~K.~Olson, M.~H.~Sarma, and M.~Sundaralingam (Adenine Press, Schenectady, 1987).

\bibitem{nelson} P.~C.~Nelson, Phys.\ Rev.\ Lett.\ \textbf{80}, 5810 (1998).

\bibitem{bensimon} D.~Bensimon, D.~Dohmi, and M.~Mezard, Europhys.\ Let.\ \textbf{42}, 97 (1998).

\bibitem{wiggins2} P.~A.~Wiggins, private communication.

\bibitem{munteanu} M.~G.~Munteanu, K.~Vlahovicek, S.~Parthasarathy, I.~Simon, and S.~Pongor, TIBS \textbf{23}, 341 (1998).

\bibitem{tkachenko} A.~V.~Tkachenko, to be published, \texttt{arXiv.org: q-bio.BM/0703026}.

\bibitem{vologodskaia} M.~Vologodskaia and A.~Vologodskii, J.\ Mol.\ Biol.\ \textbf{317}, 205 (2002).

\bibitem{molbec} G.~Bonnet, S.~Tyagi, A.~Libchaber, and F.~R.~Kramer, Proc.\ Natl.\ Acad.\ Sci.\ U.S.A.\ \textbf{96}, 6171 (1999).

\end{thebibliography}
\end{document}